\documentclass[letterpaper]{article}
\usepackage{tikz}
\usetikzlibrary{shadings}
\usepackage{aaai}
\usepackage{times}
\usepackage{graphicx}
\usepackage{helvet}
\usepackage[utf8]{inputenc}
\usepackage{courier}
\usepackage{balance}
\usepackage{makecell}
\usepackage{booktabs}
\usepackage{array,colortbl,multirow,multicol,booktabs,ctable}
\usetikzlibrary{shadings}

\newcommand{\wagner}[1]{{}}
\newcommand{\pedro}[1]{{}}
\newcommand{\manoel}[1]{{}}

\begin{document}

\title{Characterizing and Detecting Hateful Users on Twitter\thanks{This is an extended version of the homonymous short paper to be presented at ICWSM-18.}}
\author{Manoel Horta Ribeiro, Pedro H. Calais,  Yuri A. Santos, Virgílio A. F. Almeida, Wagner Meira Jr.\\ 
\texttt{\{manoelribeiro,pcalais,yuriasantos,virgilio,meira\}@dcc.ufmg.br} \\
Universidade Federal de Minas Gerais\\ Belo Horizonte, Minas Gerais, Brazil}
\nocopyright
\maketitle

\begin{abstract}
Most current approaches to characterize and detect hate speech focus on \textit{content} posted in Online Social Networks.
They face shortcomings to collect and annotate hateful speech due to the incompleteness and noisiness of OSN text and the subjectivity of hate speech. 
These limitations are often aided with constraints that oversimplify the problem, such as considering only tweets containing hate-related words.
In this work we partially address these issues by shifting the focus towards \textit{users}.
We develop and employ a robust methodology to collect and annotate hateful users which does not 
depend directly on lexicon and where the users are annotated given their entire profile. 
This results in a sample of Twitter's retweet graph containing $100,386$ users, out of which $4,972$ were annotated. 
We also collect the users who were banned in the three months that followed the data collection.
We show that hateful users differ from normal ones in terms of their activity patterns, word usage and as well as network structure.
We obtain similar results comparing the neighbors of hateful vs. neighbors of normal users and also suspended users vs. active users, 
increasing the robustness of our analysis.
We observe that hateful users are densely connected, and thus formulate the hate speech detection problem as a task of semi-supervised learning over a graph, exploiting the network of connections on Twitter. 
We find that a node embedding algorithm, which exploits the graph structure, outperforms content-based approaches for the detection of both hateful ($95\%$ AUC vs $88\%$ AUC) and suspended users ($93\%$ AUC vs $88\%$ AUC).
Altogether, we present a user-centric view of hate speech, paving the way for better detection and understanding of this relevant and challenging issue.
\end{abstract}

\section{Introduction} 
\label{sec:introduction}

The importance of understanding hate speech  in Online Social Networks (OSNs) is manifold.
Countries such as Germany have strict legislation against the practice~\cite{stein1986history}, the presence of such content may pose problems for advertisers~\cite{youtubeboycott} and users~\cite{sabatini2017online}, and manually inspecting all possibly hateful content in OSNs is unfeasible~\cite{schmidt2017survey}.
Furthermore, the trade-off between banning such behavior from platforms and not censoring dissenting opinions is a major societal issue~\cite{rainie2017future}.

This scenario has motivated work that aims to understand and detect hateful content~\cite{greevy2004classifying,warner2012detecting,burnap2016us}, by 
creating representations for tweets or comments in an OSN, \textit{e.g.}~word2vec~\cite{mikolov2013efficient}, and then classifying them as hateful or not, often drawing insights on the nature of hateful speech. 
However, in OSNs, the meaning of such content is often not self-contained, referring, for instance, to some event which just happened and 
the texts are packed with informal language, spelling errors, special characters and sarcasm~\cite{dhingra2016tweet2vec,riloff2013sarcasm}.
Besides that, hate speech itself is highly subjective, reliant on temporal, social and historical context, and occurs sparsely~\cite{schmidt2017survey}.
These problems, although observed, remain unaddressed~\cite{davidson2017automated,magu2017detecting}. Consider the tweet:

\begin{quote}
\textit{Timesup, yall getting w should have happened long ago}
\end{quote}

\noindent
Which was in reply to another tweet that mentioned the holocaust. Although the tweet, whose author's profile contained white-supremacy imagery, incited violence, it is hard to conceive how this could be detected as hateful with only textual features. Furthermore, the lack of hate-related words makes it difficult for this kind of tweet to be sampled. 

Fortunately, as we just hinted, the data in posts, tweets or messages are not the only signals we may use to study hate speech in OSNs.
Most often, these signals are linked to a profile representing a person or organization. 
Characterizing and detecting hateful \emph{users} shares much of the benefits of detecting hateful content and presents plenty of opportunities to explore a richer feature space.
Furthermore, on a practical hate speech guideline enforcement process, containing humans in the loop, its is natural that user profiles will be checked, rather than isolated tweets. The case can be made that this wider context is sometimes \textit{needed} to define hate speech, such as in the example, where the abuse was made clear by the neo-nazi signs in the user's profile.

Analyzing hate on a \textit{user-level} rather than \textit{content-level} enables our characterization to explore not only content, but also dimensions such as the user's activity and connections.
Moreover, it allows us to use the very structure of Twitter's network in the task of detecting hateful users~\cite{hamilton2017representation}.

\begin{figure}[ht]
\centering
\includegraphics[width=0.66\linewidth]{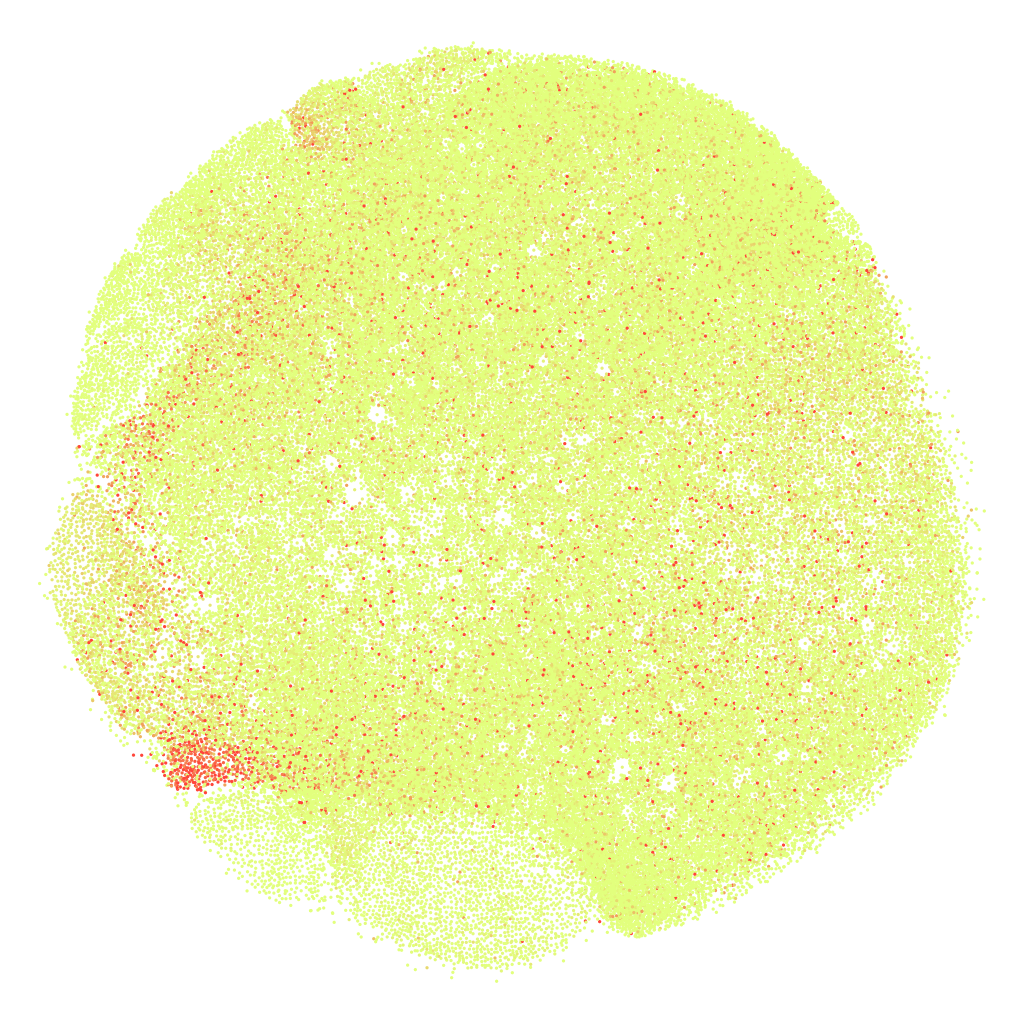}
\caption{Network of $100,386$ users sampled from Twitter after our diffusion process. Red nodes indicate the proximity of users to those who employed words in our lexicon.}
\label{fig:hintwitter}
\end{figure}

\noindent
\subsubsection{Present Work}
In this paper we  characterize and detect hateful \textit{users} on Twitter, which we define according to Twitter's hateful conduct guidelines.
We collect a dataset of $100,386$ users along with up to $200$ tweets for each with a random-walk-based crawler on Twitter's retweet graph.
We identify users that employed words from a set of hate speech related lexicon, and generate a subsample selecting users that are in different distances to such users. These are manually annotated as hateful or not through crowdsourcing.
The aforementioned distances are real valued numbers obtained through a diffusion process in which the users who used the words in the lexicon are seeds.
We create a dataset containing $4,972$ manually annotated users, of which $544$ were labeled as hateful.
We also find the users that have been suspended after the data collection - before and after Twitter's guideline changes, which happened on the 18/Dec/17.
\pedro{real-valued}

Studying these users, we find significant differences between the activity patterns of hateful and normal users: hateful users tweet more frequently, follow more people each day and their accounts are more short-lived and recent

While the media stereotypes hateful individuals as ``lone wolves''~\cite{lonewolf}, we find that hateful users are not in the periphery of the retweet network we sampled. 
Although they have less followers, the median for several network centrality measures in the retweet network is higher for those users. 
We also find that these users do not seem to behave like spammers.

A lexical analysis using \textit{Empath}~\cite{fast2016empath} shows that their choice of vocabulary is different:  words related to hate, anger and politics occur \textit{less} often when compared to their normal counterparts, and words related to masculinity, love and curses occur more often. 
This is noteworthy, as much of the previous work directly employs hate-related words as a data-collection mechanism.

We compare the neighborhood of hateful with the neighborhood of normal users in the retweet graph, as well as accounts that have been suspended with those who were not.
We argue that these suspended accounts and accounts that retweeted hateful users are also proxies for hateful speech online, and the similar results found in many of the analyses performed increase the robustness of our findings.

We also compare users who have been banned before and after Twitter's recent guideline change, finding an increase in the number of users banned per day, but little difference in terms of their vocabulary, activity and network structure.

Finally, we find that hateful users and suspended users are very densely connected in the retweet network we sampled. Hateful users are $71$ times more likely to retweet other hateful users and suspended users are $11$ times more likely to retweet other suspended users. 
This motivates us to pose the problem of detecting hate speech as a task of supervised learning over graphs. We employ a node embedding algorithm that creates a low-dimensional representation of nodes in a network to then classify it.
We demonstrate  robust performance to detect both hateful and suspended users in such fashion ($95\%$ AUC and $93\%$ AUC) and show that this approach outperforms traditional state-of-the-art classifiers ($88\%$ AUC and $88\%$ AUC, respectively).

Altogether, this work presents a user-centric view of the problem of hate speech. Our code and data are available~\footnote{https://github.com/manoelhortaribeiro/HatefulUsersTwitter}.

\section{Background} 
\label{sec:background}
\subsubsection{Hateful Users}
We define ``hateful user'' and  ``hate speech'' according to Twitter’s guidelines. For the purposes of this paper, ``hate speech'' is any type of content that `promotes violence against or directly attack or threaten other people on the basis of race, ethnicity, national origin, sexual orientation, gender, gender identity, religious affiliation, age, disability, or disease.''~\cite{twitterguidelines} On the other hand, ``hateful user'' is a user that, according to annotators, endorses such type of content.

\subsubsection{Retweet Graph}
The retweet graph $G$ is a directed graph $G=(V,E)$ where each node $u \in V$ represents a user in Twitter, and each edge $(u_1,u_2)\in E$ implies that the user $u_1$ has retweeted $u_2$. Previous work suggests that retweets are better than followers to judge users' influence~\cite{cha2010measuring}. As influence flows in the opposite direction of retweets, we invert the graph's edges. 

\subsubsection{Offensive Language}
We employ ~\citeauthor{waseem2017understanding} definition of explicit abusive language, which defines it as \textit{language that is unambiguous in its potential to be abusive, for example language that contains racial or homophobic slurs}. The use of this kind of language doesn't imply hate speech, although there is a clear correlation~\cite{davidson2017automated}.

\subsubsection{Suspended Accounts}
Most Twitter accounts are suspended due to spam, however they are harder to reach in the retweet graph as they rarely get retweeted. 
We use Twitter's API to find the accounts that have been suspended among the $100,386$ collected users, and use these as another source for potentially hateful behavior. We collect accounts that have been suspended two months after the data collection, on 12/Dec/2017, and after Twitter's hateful conduct guideline changes, on 14/Jan/2018. The new guidelines are allegedly stricter, considering for instance, off-the-platform behavior.

\section{Data Collection} 
\label{sec:data_col}
Most previous work on detecting hate speech on Twitter employs a lexicon-based data collection, which involves sampling tweets that contain specific words~\cite{davidson2017automated,waseem2016hateful}, such as \texttt{wetb*cks} or \texttt{fagg*t}.
However, this methodology is biased towards a very direct, textual and offensive hate speech. 
It presents difficulties with statements that subtly disseminate hate with no offensive words, as in
\texttt{"Who convinced Muslim girls they were pretty?"}~\cite{davidson2017automated};
And also with the usage of code words, as in the use of the word \texttt{"skypes"}, employed to reference jews~\cite{magu2017detecting,operationgoogle};
In this scenario, we propose collecting users rather than tweets, relying on lexicon only \textit{indirectly}, and collecting the structure of these users in the social network, which we will later use to characterize and detect hate.

We represent the connections among users in Twitter using the retweet network~\cite{cha2010measuring}.
Sampling the retweet network is hard as we can only observe out-coming edges (due to API limitations), 
and as it is known that any unbiased in-degree estimation is impossible without sampling most of these ``hidden'' edges in the graph~\cite{ribeiro2012sampling}.
Acknowledging this limitation, we employ Ribeiro et al. Direct Unbiased Random Walk algorithm, which  estimates out-degrees distribution efficiently by performing random jumps in an undirected graph it constructs online~\cite{ribeiro2010estimating}. 
Fortunately, in the retweet graph the outcoming edges of each user represent the other users she - usually~\cite{guerra2017antagonism} - endorses.
With this strategy, we collect a sample of Twitter retweet graph with $100,386$ users and $2,286,592$ retweet edges along with the $200$ most recent tweets for each users, as shown in Figure~\ref{fig:hintwitter}. This graph is unbiased w.r.t. the out degree distribution of nodes.

\begin{figure}[t]
\centering
\includegraphics[width=0.85\linewidth]{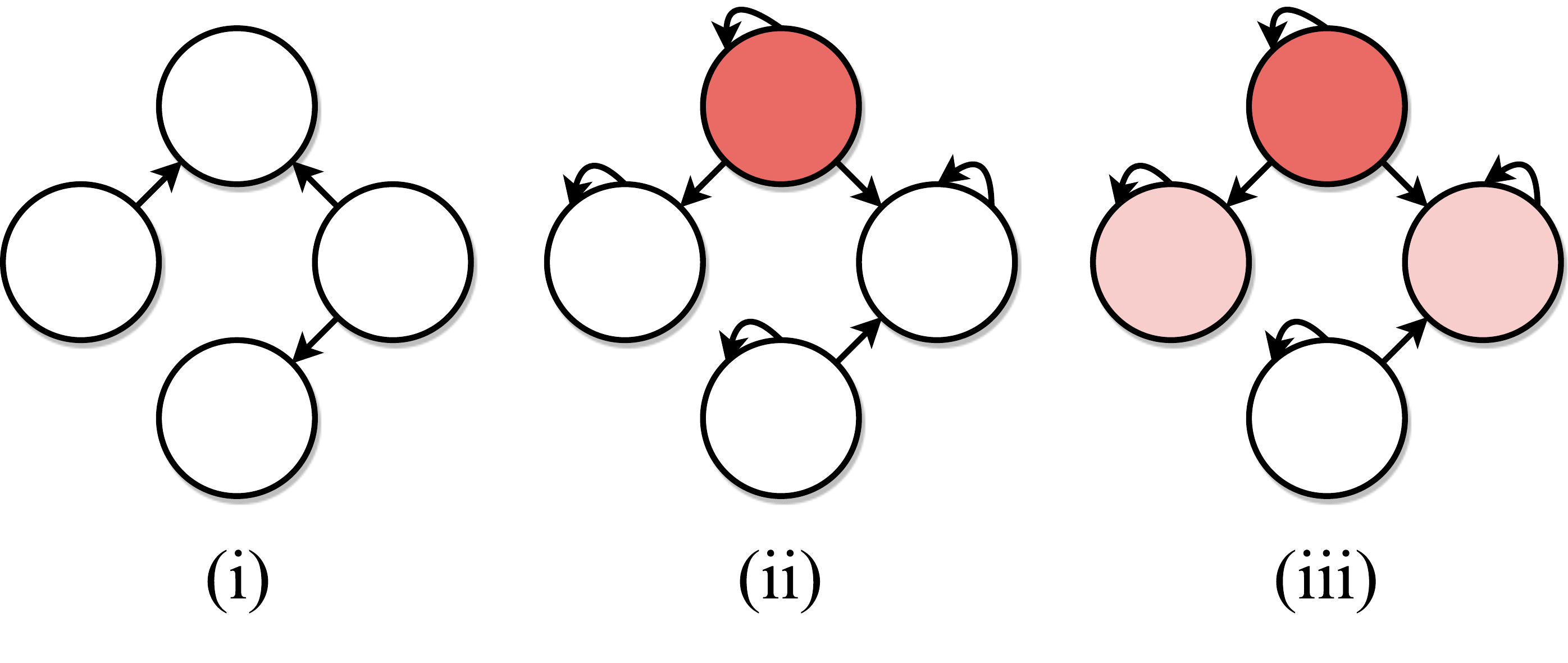}
\caption{Toy exampĺe of the diffusion process. \textit{(i)}~We begin with the sampled retweet graph $G$; \textit{(ii)} We revert the direction of the edges (the way influence flows), add self loops to every node, and mark the users who employed words in our lexicon; \textit{(iii)} We iteratively update the belief of other nodes.}
\label{fig:all_users_nets}
\end{figure}

As the sampled graph is too large to be annotated entirely, we need to select a subsample to be annotated.
If we choose tweets uniformly at random, we risk having a very insignificant percentage of hate speech in the subsample. 
On the other hand, if we choose only tweets that use obvious hate speech features, such as offensive racial slurs, we will stumble in the same problems pointed in previous work. 
We propose a method between these two extremes.
We:

\begin{enumerate}
\item Create a lexicon of words that are mostly used in the context of hate speech. This is unlike other work~\cite{davidson2017automated} as we do not consider words that are employed in a hateful context but often used in other contexts in a harmless way (\textit{e.g.} \texttt{n*gger}); We use $23$ words such as \texttt{holohoax}, \texttt{racial treason} and \texttt{white genocide}, handpicked from Hatebase.org~\cite{hatebase}, and ADL's hate symbol database~\cite{adlhate}. 
\item Run a diffusion process on the graph based on DeGroot's Learning Model~\cite{golub2010naive}, assigning an initial belief $p_{i}^{0} = 1$ to each user $u_i$ who employed the words in the lexicon; This prevents our sample from being excessively small or biased towards some vocabulary.
\item Divide the users in $4$ strata according to their associated beliefs after the diffusion process, and perform a stratified sampling, obtaining up to $1500$ user per strata.
\end{enumerate}

\begin{figure*}[ht]
\centering
\includegraphics[width=.90\textwidth]{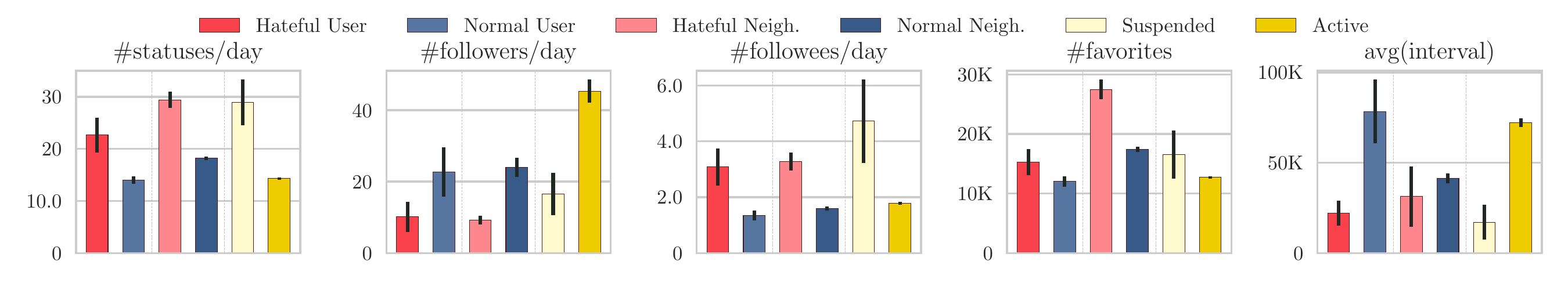}
\caption{Average values for several activity-related statistics for hateful users, normal users, users in the neighborhood of those, and suspended/active users. The \texttt{avg(interval)} was calculated on the $200$ tweets extracted for each user. Error bars represent $95\%$ confidence intervals. The legend used in this graph is kept in the remainder of the paper.}
\label{fig:attributes}
\end{figure*}

\begin{figure}[t]
\centering
\includegraphics[width=\linewidth]{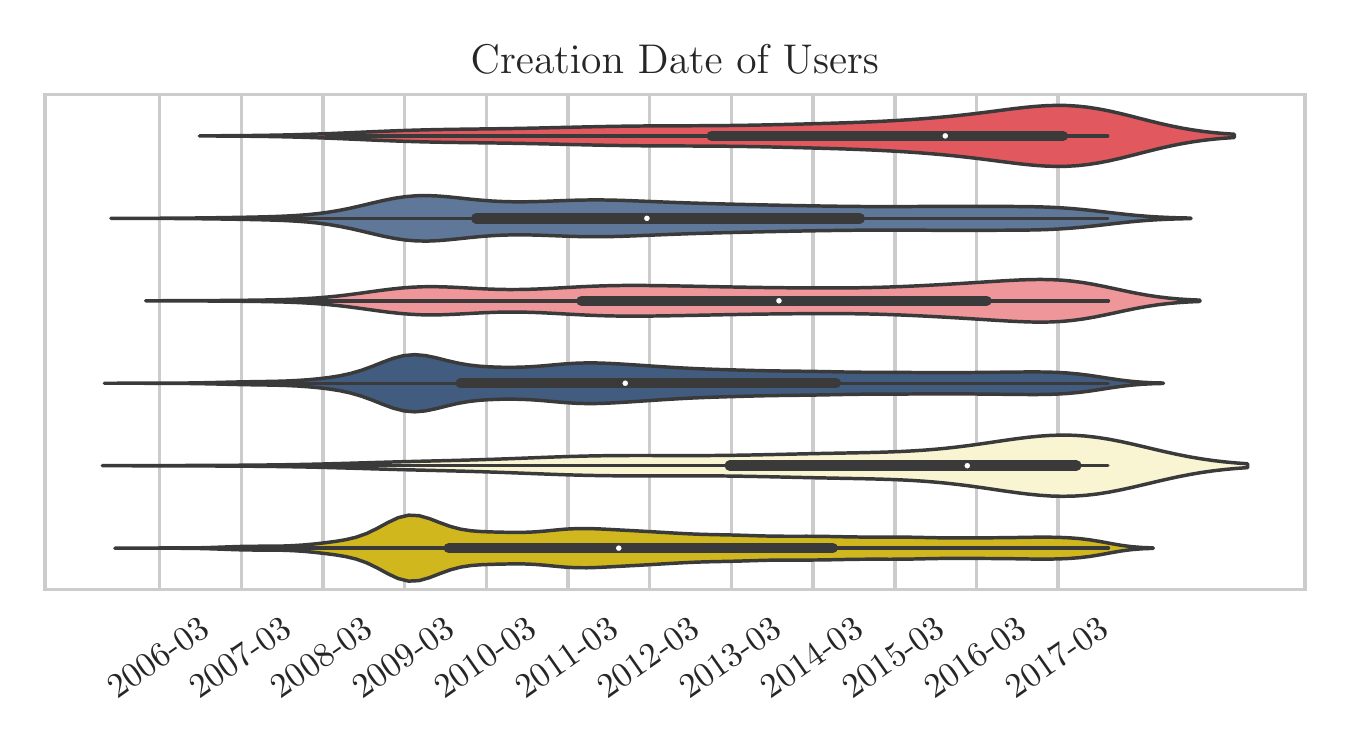}
\caption{KDEs of the creation dates of user accounts. The white dot indicates the median and the thicker bar indicates the first and third quartiles.}
\label{fig:created_at}
\end{figure}

We briefly present our diffusion model, as illustrated in Figure~\ref{fig:all_users_nets}. 
Let $A$ be the adjacency matrix of our retweeted graph $G=(V,E)$ where each node $u \in V$ represents a user and each edge $(u,v)\in E$ represents a retweet. We have that $A[u,v] = 1$ if $u$ retweeted $v$. 
We create a transition matrix $T$ by inverting the edges in $A$ (as influence flows from the retweeted user to the user who retweeted him or her), adding a self loop to each of the nodes and then normalizing each row in $A$ so it sums to $1$. This means each user is equally influenced by every user he or she retweets.

We then associate a belief $p_{i}^{0} = 1$ to every user who employed one of the words in our lexicon, and  $p_{i}^{0} = 0$ to all who didn't. 
Lastly, we create new beliefs $\mathbf{p}^{t}$ using the update rule: $\mathbf{p}^{t} = T\mathbf{p}^{t-1}$.
Notice that all the beliefs $p_{i}^{t}$ converge to the same value as $t \rightarrow \infty$, thus we run the diffusion process with $t=2$. 
With this real value ($p_{i}^{2} \in [0,1]$) associated with each user, we get 4 strata by randomly selecting up to $1500$ users with $p_{i}$ in the intervals $[0,.25)$, $[.25,.50)$, $[.50,.75)$ and $[.75,1]$.
This ensures that we annotate users that didn't employ any of the words in our lexicon, yet have a high potential to be hateful due to homophily.

We annotate $4,972$ users as hateful or not using \textit{CrowdFlower}, a crowdsourcing service. The annotators were given the definition of hateful conduct according to Twitter's guidelines and asked, for each user:

\begin{quote}
\textit{Does this account endorse content that is humiliating, derogatory or insulting towards some group of individuals (gender, religion, race) or support narratives associated with hate groups (white genocide, holocaust denial, jewish conspiracy, racial superiority)?}
\end{quote}

Annotators were asked to consider the entire profile (limiting the tweets to the ones collected) rather than individual publications or isolate words and were given examples of terms and codewords in ADL's hate symbol database. Each user profile was independently annotated by $3$ annotators, and, if there was disagreement, up to $5$ annotators. 
In the end,  $544$ hateful users and $4,427$ 
 normal ones were identified by them.
The sample of the retweet network was collected between the 1st and 7th of Oct/17, and annotation began immediately after.
We also obtained all users suspended up to 12/Dec/17 ($387$) and up to 14/Jan/18 ($668$).

\section{Characterizing Hateful Users} 
\label{sec:charac}
We analyze how hateful and normal users differ w.r.t. their activity, vocabulary and network centrality.
We also compare the neighbors of hateful and of normal users, and suspended/active users to reinforce our findings, as homophily suggests that the neighbors will share a lot of characteristics with annotated users, and as suspended users may have been banned because of hateful conduct \footnote{We use suspended and banned interchangeably.}. 
We compare those in pairs as the sampling mechanism for each of the populations is different.
We argue that each one of these pairs contains a proxy for hateful speech in Twitter, and thus inspecting the three increases the robustness of our analysis. 
P-values given are from unequal variances t-tests to compare the averages across distinct populations.
When we refer to ``hateful users'', we refer to the ones annotated as hateful. 
The number of users in each of these groups is given in the table bellow: 

\begin{table}[h]
\centering
\caption{Number of users in each group.}
\label{tab:numbers}
\begin{tabular}{@{}cc|cc|cc@{}}
\toprule
 {\small Hateful} & 
 {\small Normal}  & 
 \makecell{{\small Hateful} \\ {\small Neighbors}} & 
 \makecell{{\small Normal} \\ {\small Neighbors}} & 
 {\small Banned} & 
 {\small Active}
  \\ \midrule
 $544$ & $4427$ & $3471$ & $33564$ & $668$ & $99718$
 \\ \bottomrule
\end{tabular}
\end{table}

\subsection{Hateful users have newer accounts} 
The account creation date of users is depicted in Figure~\ref{fig:created_at}. 
Hateful users were created later than normal ones (p-value $< 0.001$). 
A hypothesis for this difference is that hateful users are banned more often due to Twitter's guidelines infringement. 
This resonates with existing methods for detecting fake accounts in which using the account's creation date have been successful~\cite{viswanath2015strength}. 
We obtain similar results w.r.t. the 1-neighborhood of such users, where the hateful neighbors were also created more recently (p-value $< 0.001$), and also when comparing suspended and active accounts (p-value $< 0.001$).

\begin{figure}[t]
\centering
\includegraphics[width=\linewidth]{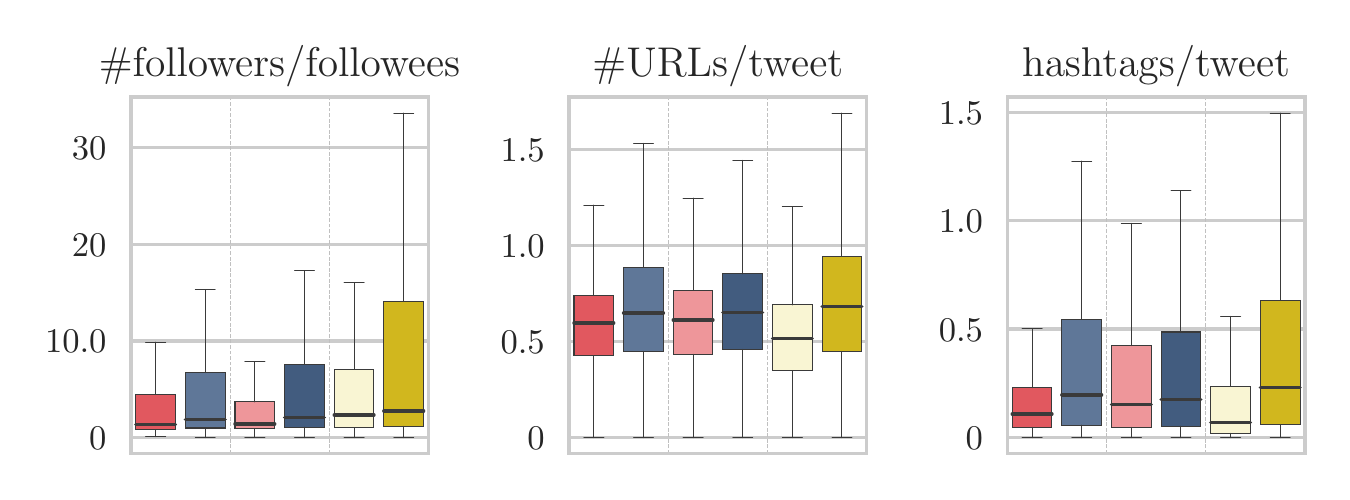}
\caption{Boxplots for the distribution of metrics that indicate spammers. Hateful users have slightly \textit{less} followers per followee, \textit{less} URLs per tweet, and \textit{less} hashtags per tweet.}
\label{fig:spam}
\end{figure}

\begin{figure*}[t]
\centering
\includegraphics[width=\textwidth]{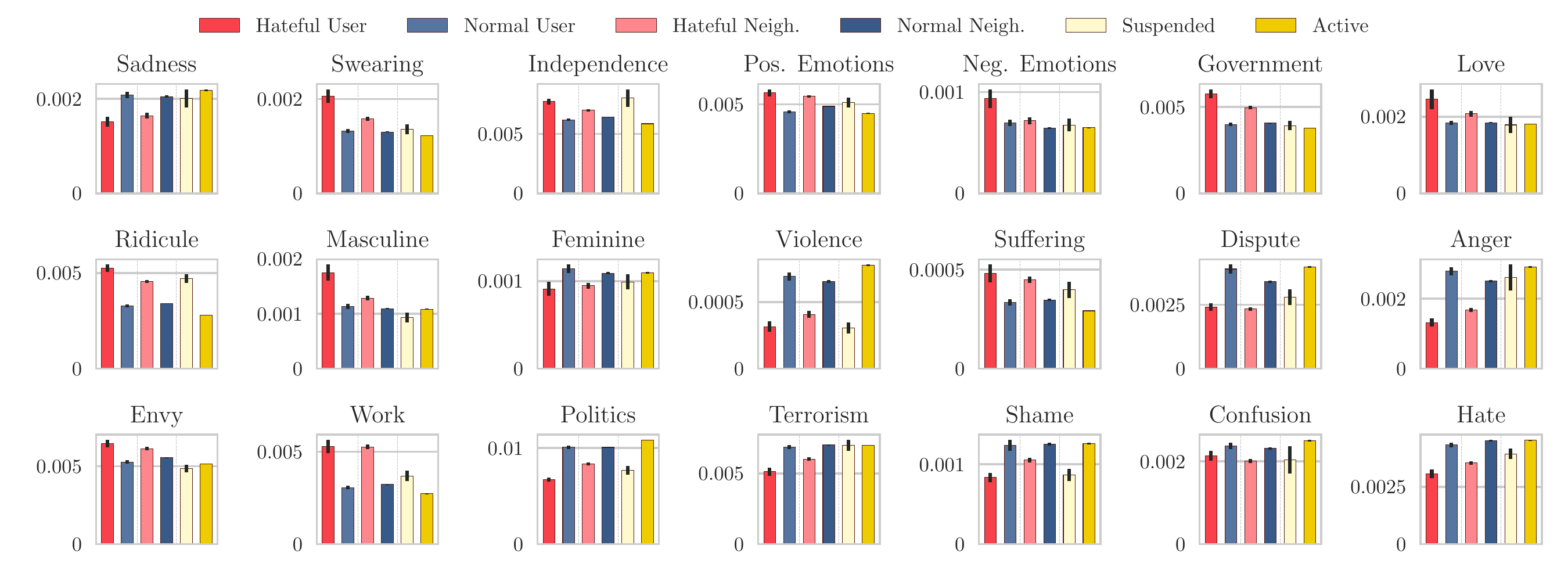}
\caption{Average values for the relative occurrence of several categories in \textit{Empath}. Notice that not all Empath categories were analyzed and that the to-be-analyzed categories were chosen before-hand to avoid spurious correlations. Error bars represent $95\%$ confidence intervals.}
\label{fig:lex}
\end{figure*}

\begin{figure}[t]
\centering
\includegraphics[width=\linewidth]{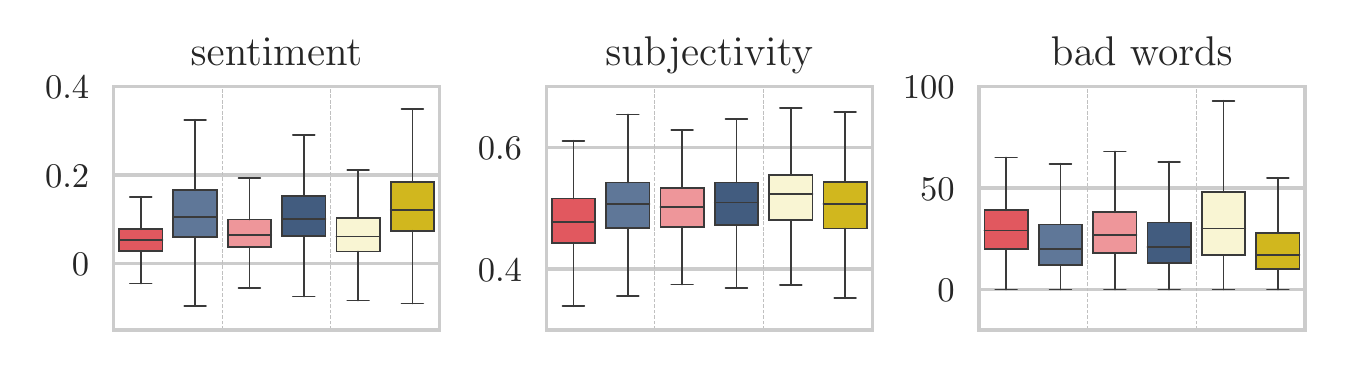}
\caption{Box plots for the distribution of sentiment and subjectivity and bad-words usage. Suspended users, hateful users and their neighborhood are more negative, and use more bad words than their counterparts.}
\label{fig:sent}
\end{figure}

\begin{figure}[t]
\centering
\includegraphics[width=\linewidth]{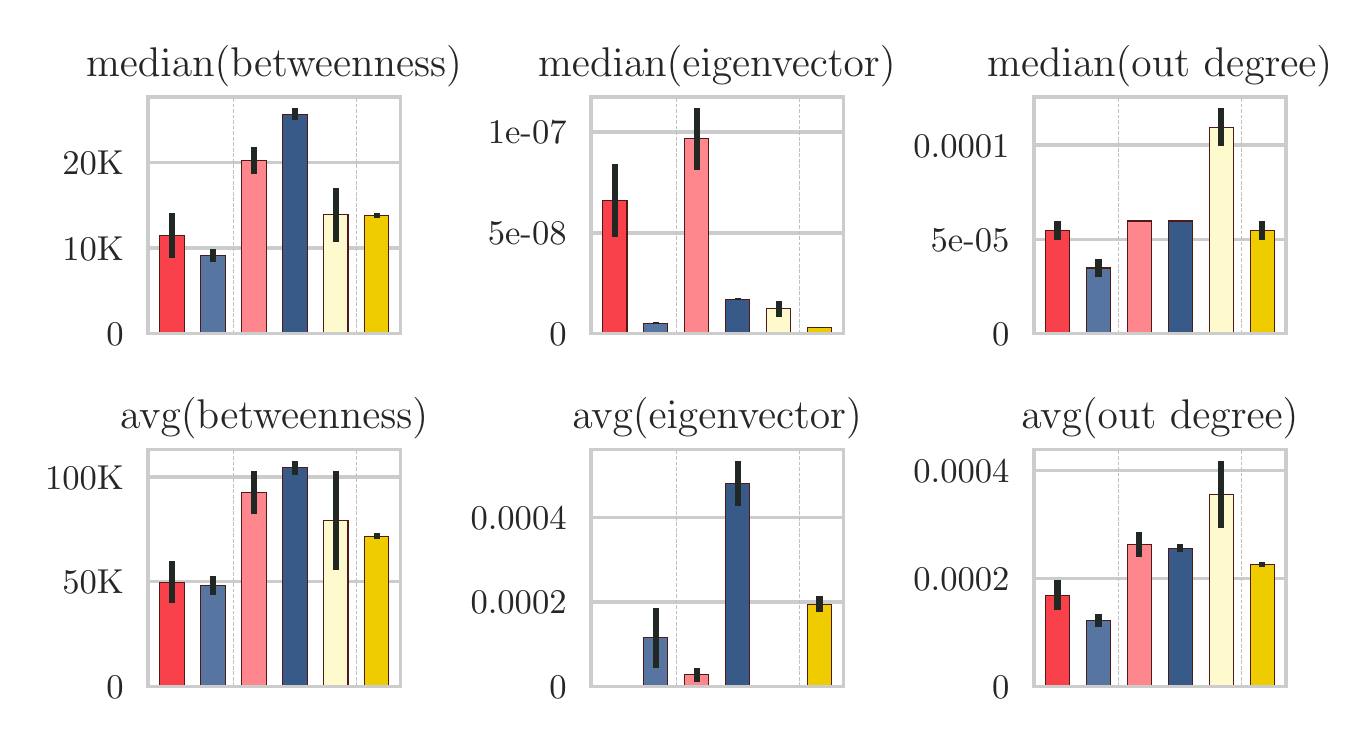}
\caption{Network centrality metrics for hateful and normal users, their neighborhood, and suspended/non-suspended users calculated on the sampled retweet graph.}
\label{fig:betweenness}
\end{figure}

\subsection{Hateful users are power users} 
Other interesting metrics for analysis are the number of tweets, followers, followees and favorite tweets a user has, and the interval in seconds between their tweets. 
We show these statistics in Figure~\ref{fig:attributes}. 
We normalize the number of tweets, followers and followees by the number of days the users have since their account creation date. 
Our results suggest that hateful users are ``power users'' in the sense that they tweet more, in shorter intervals, favorite more tweets by other people and follow other users more (p-values $<0.01$).
The analysis yields similar results when we compare the 1-neighborhood of hateful and normal users, and when comparing suspended and active accounts (p-values $<0.01$, except for the number of favorites when comparing suspended/active users, and for the average interval, when comparing the neighborhood).

\subsection{Hateful users don't behave like spammers}
We investigate whether users that propagate hate speech are spammers.
We analyze metrics that have been used by previous work to detect spammers, such as the numbers of URLs per tweet, of hashtags per tweet and of followers per followees~\cite{benevenuto2010detecting}. 
The boxplot of these distributions is shown on Figure~\ref{fig:spam}. 
We find that hateful users use, in average, \textit{less} hashtags (p-value $< 0.001$) and \textit{less} URLs (p-value $< 0.001$) per tweet than normal users. 
The same analysis holds if we compare the 1-neighborhood of hateful and non-hateful, or suspended and active users (with p-values $< 0.05$, except for the number of followers per followees, where there is no statistical significance to the t-test). 
Additionally, we also find that, in average, normal users have more followers per followees than hateful ones (p-value $< 0.05$), which also happens for their neighborhood (p-value $< 0.05$). 
This suggests that the hateful and suspended users do not use systematic and programmatic methods to deliver their content.
Notice that it is not possible to extrapolate this finding to Twitter in general, as there maybe be hateful users with other behaviors which our data collection methodology does not consider, as we do not specifically look for trending topics or popular hashtags.

\subsection{The median hateful user is more central}
We analyze different measures of centrality for users, as depicted in Figure~\ref{fig:betweenness}. 
The median hateful user is more central in all measures when compared to their normal counterparts. 
This is a counter-intuitive finding, as hateful crimes have long been associated with ``lone wolves'', and anti-social people~\cite{lonewolf}.
We observe similar results when comparing the median eigenvector centrality of the neighbors of hateful and normal users, as well as suspended and active users. In the latter pair, suspended users also have higher median out degree.
When analyzing the average for such statistics, we observe the average eigenvector centrality is higher for the opposite sides of the previous comparisons. This happens because some very influential users distort the value: for example, the  $970$ most central  users according to the metric are normal.
Notice that despite of this, hateful and suspended users have higher average out degree than normal and active users respectively (p-value  $< 0.05$).

\subsection{Hateful users use non-trivial vocabulary} 
We characterize users w.r.t. their content  with \textit{Empath}~\cite{fast2016empath}, as depicted in Figure~\ref{fig:lex}.
Hateful users use \textit{less} words related to hate, anger, shame and terrorism, violence, and sadness when compared to normal users (with p-values $< 0.001$). 
A question this raises is how sampling tweets based exclusively in a hate-related lexicon biases the sample of content to be annotated to a very specific type of ``hate-spreading'' user, and
reinforces the claims that sarcasm, code-words and very specific slang plays a significant role in defining such users~\cite{davidson2017automated,magu2017detecting}.

Categories of words more used by hateful users include positive emotions, negative emotions, suffering, work, love and swearing (with p-values $< 0.001$), suggesting the use of emotional vocabulary.
An interesting direction would be to analyze the sensationalism of their statements, as it has been done in the context of \textit{clickbaits}~\cite{chen2015misleading}.
When we compare the neighborhood of hateful and normal users and suspended vs active users, we obtain very similar results (with p-values $< 0.001$ except for when comparing suspended vs. active users usage of anger, terrorism and sadness, swearing and love).
Overall, the non-triviality of the vocabulary of these groups of users reinforces the difficulties found in the NLP approaches to sample, annotate and detect hate speech~\cite{davidson2017automated,magu2017detecting}.

\begin{figure}[t]
\centering
\includegraphics[width=1\linewidth]{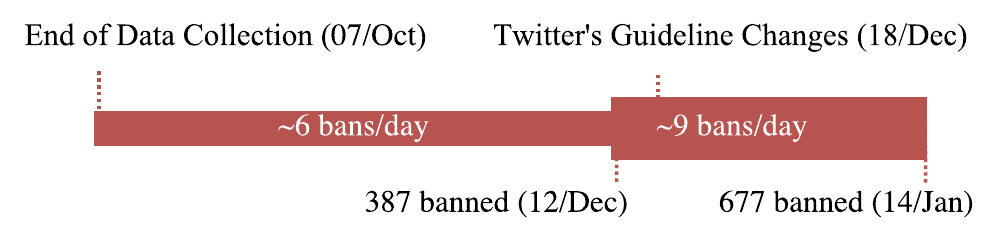}
\caption{Corhort-like depiction of the banning of users.}
\label{fig:cohor_hate}
\end{figure}

\begin{table}[t]
\centering
\caption{Percentage/number of accounts that got suspended up before and after the guidelines changed.}
\label{tab:sus}
\begin{tabular}{@{}l|lll@{}}
\toprule
Susp. Accounts & Hateful   & Normal    & Others           \\ \midrule
2017-12-12                             & $$9.09\%/55$$  & $$0.32\%/14$$ & $$0.33\%/318$$ \\ \midrule
2018-01-14                             & $$17.64\%/96$$ & $$0.90\%/40$$ & $$0.55\%/532$$ \\ \bottomrule
\end{tabular}
\end{table}

We also explore the sentiment in the tweets users write using a corpus based approach, as depicted in Figure~\ref{fig:sent}. We find  that sentences written by hateful and suspended users are more negative, and are less subjective (p-value $<0.001$). The neighbors of hateful users in the retweet graph are also more negative (p-value $<0.001$), however not less subjective. We also analyze the distribution of profanity per tweet in hateful and non-hateful users. The latter is obtained by matching all words in Shutterstock's ``List of Dirty, Naughty, Obscene, and Otherwise Bad Words''~\footnote{{\tiny https://github.com/LDNOOBW/List-of-Dirty-Naughty-Obscene-and-Otherwise-Bad-Words}}. We find that suspended users, hateful users and their neighbors employ more profane words per tweet, also confirming the results from the analysis with \textit{Empath} (p-value $< 0.01$). 

\begin{table}[t]
\centering
\caption{Occurrence of the edges between hateful 
\tikz{\definecolor{tempcolor}{RGB}{248,65,74} \draw[fill=tempcolor,line width=1pt]  circle(0.8ex);} 
and normal
\tikz{\definecolor{tempcolor}{RGB}{86,118,161} \draw[fill=tempcolor,line width=.5pt]  circle(0.8ex);} 
users,
and between suspended
\tikz{\definecolor{tempcolor}{RGB}{255,250,205} \draw[fill=tempcolor,line width=1pt]  circle(0.8ex);} 
and active
\tikz{\definecolor{tempcolor}{RGB}{239,204,0} \draw[fill=tempcolor,line width=.5pt]  circle(0.8ex);} 
users. Results are normalized w.r.t. to the type of the source node, as in: $P($source type$\rightarrow$dest type$|$source type$)$. Notice that the probabilities do not add to $1$ in hateful and normal users as we don't present the statistics for non-annotated users.}
\label{tab:links}
\begin{tabular}{@{}cc|cc@{}}
\toprule
Node Type & ($\%$) & Node Type  & ($\%$) \\ \midrule
\tikz{\definecolor{tempcolor}{RGB}{248,65,74} \draw[fill=tempcolor,line width=1pt]  circle(0.8ex);} 
$\rightarrow$
\tikz{\definecolor{tempcolor}{RGB}{248,65,74} \draw[fill=tempcolor,line width=1pt]  circle(0.8ex);} 
&
$41.50$
& 
\tikz{\definecolor{tempcolor}{RGB}{248,65,74} \draw[fill=tempcolor,line width=1pt]  circle(0.8ex);} 
$\rightarrow$
\tikz{\definecolor{tempcolor}{RGB}{86,118,161} \draw[fill=tempcolor,line width=.5pt]  circle(0.8ex);} 
&
$13.10$
\\ \midrule
\tikz{\definecolor{tempcolor}{RGB}{86,118,161} \draw[fill=tempcolor,line width=.5pt]  circle(0.8ex);} 
$\rightarrow$
\tikz{\definecolor{tempcolor}{RGB}{86,118,161} \draw[fill=tempcolor,line width=.5pt]  circle(0.8ex);} 
&
$15.90$
& 
\tikz{\definecolor{tempcolor}{RGB}{86,118,161} \draw[fill=tempcolor,line width=.5pt]  circle(0.8ex);} 
$\rightarrow$
\tikz{\definecolor{tempcolor}{RGB}{248,65,74} \draw[fill=tempcolor,line width=1pt]  circle(0.8ex);} 
&
$2.86$\\ \midrule
\tikz{\definecolor{tempcolor}{RGB}{255,250,205} \draw[fill=tempcolor,line width=1pt]  circle(0.8ex);} 
$\rightarrow$
\tikz{\definecolor{tempcolor}{RGB}{255,250,205} \draw[fill=tempcolor,line width=1pt]  circle(0.8ex);} 
&
$7.50$
& 
\tikz{\definecolor{tempcolor}{RGB}{255,250,205} \draw[fill=tempcolor,line width=1pt]  circle(0.8ex);} 
$\rightarrow$
\tikz{\definecolor{tempcolor}{RGB}{239,204,042} \draw[fill=tempcolor,line width=.5pt]  circle(0.8ex);} 
&
$92.50$
\\ \midrule
\tikz{\definecolor{tempcolor}{RGB}{239,204,0} \draw[fill=tempcolor,line width=.5pt]  circle(0.8ex);} 
$\rightarrow$
\tikz{\definecolor{tempcolor}{RGB}{239,204,0} \draw[fill=tempcolor,line width=.5pt]  circle(0.8ex);} 
&
$99.35$
& 
\tikz{\definecolor{tempcolor}{RGB}{239,204,0} \draw[fill=tempcolor,line width=.5pt]  circle(0.8ex);} 
$\rightarrow$
\tikz{\definecolor{tempcolor}{RGB}{255,250,205} \draw[fill=tempcolor,line width=1pt]  circle(0.8ex);} 
&
$0.65$\\ \bottomrule
\end{tabular}

\end{table}

\subsection{More users are banned after the guideline changes, but they are similar to the ones banned before} 

 Twitter has changed its enforcement of hateful conduct guidelines in 18/Dec/2017. 
We analyze the differences among accounts that have been suspended two months after the end of the annotation, in 12/Dec/2017 and in 14/Jan/2018. 

The intersection between these groups and the ones we annotated as hateful or not is shown in Table~\ref{tab:sus}. 
In the first period from the end of the data annotation to the 12/Dec, there were approximately $6.45$ banned users a day whereas in the second period there were $9.05$. This trend, illustrated in Figure~\ref{fig:cohor_hate}, suggests an increased banning activity.

Performing the lexical analysis we previously applied to compare hateful and normal users we do not find statistically significant difference w.r.t. the averages for users banned before and after the guideline change (except for government-related words, where p-value $< 0.05$). 
We also analyze the number of tweets, followers/followees, and the previously mentioned centrality measures, and observe no statistical significance in difference between the averages or the distributions (which were compared using KS-test). This suggests that Twitter has not changed the type of users banned.

\subsection{Hateful users are densely connected} 

Finally, we analyze the frequency at which hateful and normal users, as well as suspended and active users, interact within their own group and with each other. 
Table~\ref{tab:links} depicts the probability of an node of a given type retweeting other type of node. 
We find that $41\%$ of the retweets of hateful users are to other hateful users, which means that they are $71$ times more likely to retweet another hateful user, considering the occurrence of hateful users in the graph. We observe a similar phenomenon with suspended users, which have $7\%$ of their retweets redirected towards other suspended users. As suspended users correspond to only $0.68\%$ of the users sampled, this means they are approximately $11$ times more likely to retweet other suspended users.

The high density of connections among hateful and suspended users suggest a strong modularity. We exploit this, along with activity and network centrality attributes to robustly detect these users.

\begin{table*}[!ht]
\centering
\caption{Prediction results and standard deviations for the two proposed settings: detecting hateful users and detecting suspended users. The semi-supervised node embedding approach performs better than state-of-the-art supervised learning algorithms in all the assessed criteria, suggesting the benefits of exploiting the network structure to detect hateful and suspended users. }
\label{tab:res}
\begin{tabular}{cc|ccc|ccc}\toprule
 & & \multicolumn{3}{c}{Hateful/Normal} & \multicolumn{3}{c}{Suspended/Active} \\  \midrule
Model & Features & Accuracy & F1-Score & AUC & Accuracy & F1-Score & AUC
\\ \midrule
\texttt{GradBoost}   & \texttt{user+glove}  & $84.6 \pm 1.0$ & $52.0 \pm 2.2$ & $88.4 \pm 1.3$
                                            & $81.5 \pm 0.6$ & $48.4 \pm 1.1$ & $88.6 \pm 0.1$ \\ 
                     & \texttt{glove}       & $84.4 \pm 0.5$ & $52.0 \pm 1.3$ & $88.4 \pm 1.3$
                                            & $78.9 \pm 0.7$ & $44.8 \pm 0.7$ & $87.0 \pm 0.5$ \\  
\midrule
\texttt{AdaBoost}    & \texttt{user+glove}  & $69.1 \pm 2.4$ & $37.6 \pm 2.4$ & $85.5 \pm 1.4$
                                            & $70.1 \pm 0.1$ & $38.3 \pm 0.9$ & $84.3 \pm 0.5$ \\ 
                     & \texttt{glove}       & $69.1 \pm 2.5$ & $37.6 \pm 2.4$ & $85.5 \pm 1.4$
                                            & $69.7 \pm 1.0$ & $37.5 \pm 0.8$ & $82.7 \pm 0.1$ \\  
\midrule
 \texttt{GraphSage}  & \texttt{user+glove}  & $\mathbf{90.9 \pm 1.1}$ & $\mathbf{67.0 \pm 4.1}$ & $\mathbf{95.4 \pm 0.2}$
                                            & $\mathbf{84.8 \pm 0.3}$ & $\mathbf{55.8 \pm 4.0}$ & $\mathbf{93.3 \pm 1.4}$ \\ 
                     & \texttt{glove}       & $90.3 \pm 1.9$ & $65.9 \pm 6.2$ & $94.9 \pm 2.6$
                                            & $84.5 \pm 1.0$ & $54.8 \pm 1.6$ & $93.3 \pm 1.5$ \\ \bottomrule
\end{tabular}
\end{table*}

\newpage

\section{Detecting Hateful Users}  
As we consider users and their connections in the network, we can use information that is not available for models which operate on the granularity level of tweets or comments to detect hate speech.

\begin{itemize}
\item \textbf{Activity/Network:} Features such as number of statuses, followers, followees, favorites, and centrality measurements such as betweenness, eigenvector centrality and the in/out degree of each node. We refer to these as \texttt{user}.

\item \textbf{GloVe:} We also use spaCy's off-the-shelf 300-dimensional GloVe's vector~\cite{pennington2014glove} as features. We average the representation across all words in a given tweet, and subsequently, across all tweets a user has. We refer to these as \texttt{glove}.
\end{itemize}

Using these features, we compare experimentally two traditional machine learning models known to perform very well when the number of instances is not very large: 
Gradient Boosted Trees (\texttt{GradBoost}) and Adaptive Boosting (\texttt{AdaBoost}); 
and a model aimed specifically at learning in graphs, GraphSage~\cite{hamilton2017inductive}  (\texttt{GraphSage}).
Interestingly, the latter approach is semi-supervised, and allows us to use the neighborhood of the users we are classifying even though they are not labeled, exploiting the modularity between hateful and suspended users we observed. 
The algorithm creates low-dimensional embeddings for nodes, given associated features (unlike other node embeddings, such as \texttt{node2vec}~\cite{grover2016node2vec}). 
Moreover, it is inductive - which means we don't need the entire graph to run it. 
For additional information on node embeddings methods, refer to \cite{hamilton2017representation}.

The GraphSage algorithm creates embeddings for each node given that the nodes have associated features (in our case the \textit{GloVe} embeddings and activity/network-centrality attributes associated with each user). Instead of generating embeddings for all nodes, it learns a function that generate embeddings by sampling and aggregating features from a node's local neighborhood. 
This strategy exploits the structure of the graph beyond merely using the features of the neighborhood of a given node.

\subsection{Experimental Settings} 

We run the algorithms trying to detect both hateful and normal users, as annotated by the crowdsourcing service, as well as trying to detect which users got banned. 
We perform a 5-fold cross validation and report the F1-score, the accuracy and the area under the ROC curve (AUC)  for all instances.

In all approaches we accounted for the class imbalance (of approximately $1$ to $10$) in the loss function. 
We keep the same ratio of positive/negative classes in both tasks, which, in practice, means we used the $4981$ annotated users in the first setting (where approximately $11\%$ were hateful) and, in the second setting, selected $6680$ users from the graph, including the $668$ suspended users, and other $5405$ users randomly sampled from the graph.

Notice that, as we are dealing with a binary classification problem, we may control the trade-off between specificity and sensitivity by varying the positive-class threshold. 
In this work we simply pick the largest value, and report the resulting $AUC$ score - which can be interpreted as the
probability of a classifier correctly ranking a random positive case higher than a random negative case.

\subsection{Results}

The results of our experiments are shown in Table~\ref{tab:res}. 
We find that the node embedding approach using the features related to both users and the \textit{GloVe} embeddings yields the best results for all metrics in the two considered scenarios. 
The Adaptative Boosting approach yields good $AUC$ scores, but incorrectly classifies many normal users as hateful, which results in a low accuracy and F1-score.

Using the features related to users makes little difference in many settings, yielding, for example, exactly the same $AUC$, and very similar accuracy/F1-score in the Gradient Boosting models trained with the two sets of parameters.
However, the usage of the retweet network yields promising results, especially because we observe improvements in both the detection of hateful users and of suspended users, which shows the performance improvement occurs independently of our annotation process. 

\section{Related Work}  
\label{sec:related}
We review previous efforts to charactere and detect hate speech in OSNs. 
Tangent problems such as cyber-bullying and offensive language are not extensively covered; refer to \citeauthor{schmidt2017survey}.

\subsection{Characterizing Hate}

Hate speech has been characterized in websites and different Online Social Networks.
\citeauthor{gerstenfeld2003hate} analyze hateful websites characterizing their \textit{modus operandi} w.r.t. monetization, recruitment, and international appeal.
\citeauthor{chau2007mining} identified and analyzed how hate-groups organize around blogs.
\citeauthor{silva2016analyzing} matches regex-like expressions on large datasets on Twitter and Whisper to characterize the targets of hate in Online Social Networks.
\citeauthor{chatzakou2017hate} characterized users and their tweets in the specific context surrounding the \#GamerGate controversy.
More generally, abuse online also has been characterized on Community-Based Question Answering ~\cite{kayes2015social}, and in Ask.fm~\cite{van2015automatic}.

\subsection{Detecting Hate}

We briefly go through different steps carried by previous work on the task detecting hate speech, analyzing the similarities and differences to this work.

\subsubsection{Data Collection} 
Many previous studies collect data by sampling OSNs with the aid of a lexicon with terms associated with hate speech~\cite{davidson2017automated,waseem2016hateful,burnap2016us,magu2017detecting}. 
This may be succeeded by expanding the lexicon with co-occurring terms~\cite{waseem2016hateful}.
Other techniques employed include matching regular expressions~\cite{warner2012detecting} and selecting features in tweets from users known to have reproduced hate speech~\cite{kwok2013locate}.
We employ a random-walk-based methodology to obtain a generic sample of Twitter's retweet graph,
use a lexicon of hateful words to obtain a subsample of potentially hateful users 
and then select users to annotate in different ``distances'' from these users, which we obtain through a diffusion process.
This allows us not to rely directly on lexicon to obtain the sample to be annotated.

\subsubsection{Annotation} 
Human annotators are used in most previous work on hate speech detection.
This labeling may be done by researchers themselves~\cite{waseem2016hateful,kwok2013locate,djuric2015hate,magu2017detecting}, 
selected annotators~\cite{warner2012detecting,gitari2015lexicon},
or crowd-sourcing services~\cite{burnap2016us}.
Hate speech speech has been pointed out as a  difficult subject to annotate on~\cite{waseem2016you,ross2017measuring}.
\citeauthor{chatzakou2017mean} annotate tweets in sessions, clustering several tweets to help annotators get a grasp on context.
We also employ \textit{CrowdFlower} to annotate our data. 
Unlike previous work, we give annotators the entire profile of an user, instead of individual tweets. We argue this provides better context for the annotators~\cite{waseem2017understanding}.
The extent additional context improves annotation quality is a promising research direction.


\subsubsection{Features} 
Features used in previous work are almost exclusively content-related. 
The content from tweets, posts and websites has been represented as n-grams,  BoWs~\cite{waseem2016hateful,kwok2013locate,magu2017detecting,greevy2004classifying,gitari2015lexicon},
and word embeddings such as \textit{paragraph2vec}~\cite{djuric2015hate}, \textit{GloVe}~\cite{pennington2014glove} and \textit{FastText}~\cite{badjatiya2017deep}.
Other techniques used to extract features from content include POS tagging, sentiment analysis and ease of reading measures~\cite{warner2012detecting,davidson2017automated,burnap2016us,gitari2015lexicon}.
\citeauthor{waseem2016hateful} employ features not related to the content itself, using the gender and the location of the creator of the content.
We use attributes related to the user's activity, his network centrality and the content he or she produced in our characterization and detection.
In the context of detecting aggression and cyber-bullying on Twitter, \citeauthor{chatzakou2017mean} employ a similar set of features as we do.

\subsubsection{Models}  
Models used to classify these features in the existing literature include supervised classification methods such as  Naïve-Bayes~\cite{kwok2013locate},
Logistic Regression~\cite{waseem2016hateful,davidson2017automated,djuric2015hate},
Support Vector Machines~\cite{warner2012detecting,burnap2016us,magu2017detecting},
Rule-Based Classifiers~\cite{gitari2015lexicon},
Random Forests~\cite{burnap2016us},
Gradient-Boosted Decision Trees~\cite{badjatiya2017deep} and 
Deep Neural Networks~\cite{badjatiya2017deep}. 
We use Gradient-Boosted Decision Trees, Adaptative Boosting and a semi-supervised node embedding approach (GraphSage).
Our experiments shows that the latter performs significantly better. 
A possible explanation for this is because hateful users retweet other hateful users very often, which makes exploiting the network structure beneficial.

\section{Conclusion and Discussion} 
We present an approach to characterize and detect hate speech on Twitter at a user-level granularity. Our methodology differs previous efforts, which focused on isolated pieces of content, such as tweets and comments.~\cite{greevy2004classifying,warner2012detecting,burnap2016us}. 
We developed a methodology to sample Twitter which consists of obtaining a generic subgraph, finding users who employed words in a lexicon of hate-related words and running a diffusion process based on DeGroot's learning model to sample for users in the neighborhood of these users. 
We then used \textit{Crowdflower}, a crowdsourcing service to manually annotate $4,988$ users, of which $544$ ($11\%$) were considered to be hateful.
We argue that this methodology aids two existing shortcomings of existing work: it allows the researcher to balance between having a generic sample and a sample biased towards a set of words in a lexicon, and it provides annotators with realistic context, which is sometimes necessary to identify hateful speech.

Our findings shed light on how hateful users differ from normal ones with respect to their user activity patterns, network centrality measurements, and the content they produce. 
We discover that hateful users have created their accounts more recently and write more negative sentences. They use lexicon associated with categories such as hate, terrorism, violence and anger \textit{less} than normal ones, and categories of words such as love, work and masculinity \textit{more} frequently.
We also find that the median hateful user is  more central and that hateful users are densely connected in the retweet network.
The latter finding motivates the use of an inductive graph embedding approach to detect hateful users, which outperforms widely used algorithms such as Gradient Boosted Trees.
As moderation of Online Social Networks in many cases analyzes users, characterizing and detecting hate on a user-level granularity is an essential step for creating workflows where humans and machines can interact to ensure OSNs obey legislation, and to provide a better experience for the average user.

Nevertheless, our approach still has limitations that may lead to interesting future research directions.
Firstly, our characterization only considered the behavior of users on Twitter, and the same scenario in other Online Social Networks such as Instagram of Facebook may present different challenges.
Secondly, although classifying hateful users provides contextual clues that are not available when looking only at a piece of content, it is still a non-trivial task, as hateful speech is subjective, and people can disagree with what is hateful or not. 
In that sense, an interesting direction would be to try to create mechanisms of consensus, where online communities could help moderate their content in a more decentralized fashion (like Wikipedia~\cite{shi2017wisdom}).
Lastly, a research question in the context of detecting hate speech on a user-level granularity that this work fails to address is \textit{how much hateful content comes from how many users}. This is particularly important as, if we have a Pareto-like distribution where most of the hate is generated by very few users, then analyzing hateful users rather than content becomes even more attractive.

An interesting debate which may arise when shifting the focus on hate speech detection from content to users is how this can potentially blur the line between individuals and their speech. 
Twitter, for instance, implied it will consider conduct occurring ``off the platform'' in making suspension decisions~\cite{slate}.
In this scenario, approaching the hate speech detection problem as we propose could allow users to be suspended to "contextual" factors - and not for a specific piece of content he or she wrote.
However, as mentioned previously, such models can be used as a first step to detect these users, which then will be assessed by humans or other more specific methods. 

The broader question this brings is to what extent a ``black-box'' model may be used to aid in tasks such as content moderation, where this model may contain accidental or intentional bias. 
These models can be used to moderate Online Social Networks, without the supervision of a human, in which case its bias could be very damaging towards certain groups, even leading to possible suppressions of individual's human rights, notably the right to free speech.
Another option would be to make a clear distinction between using the model to detect possibly hateful or inadequate content and delegating the task of moderation exclusively to a human.
Although there are many shades of gray between these two approaches, an important research direction is how to make the automated parts of the moderation process fair, accountable and transparent, which is hard to achieve even for content-based approaches.

\section*{Acknowledgments.} 
We would like to thank Nikki Bourassa, Ryan Budish, Amar Ashar and Robert Faris from the BKC for their insightful suggestions. This work was partially supported by CNPq, CAPES and Fapemig, as well as projects InWeb, INCT-Cyber, MASWEB, BigSea and Atmosphere.
\balance
\begin{small}
\bibliographystyle{aaai}
\bibliography{bibfile.bib}

\begin{thebibliography}{}

\bibitem[\protect\citeauthoryear{ADL}{2017}]{adlhate}
ADL.
\newblock 2017.
\newblock {ADL Hate Symbols Database}.

\bibitem[\protect\citeauthoryear{Badjatiya \bgroup et al\mbox.\egroup
  }{2017}]{badjatiya2017deep}
Badjatiya, P.; Gupta, S.; Gupta, M.; and Varma, V.
\newblock 2017.
\newblock Deep learning for hate speech detection in tweets.
\newblock In {\em Proceedings of the 26th International Conference on World
  Wide Web Companion},  759--760.
\newblock International World Wide Web Conferences Steering Committee.

\bibitem[\protect\citeauthoryear{Benevenuto \bgroup et al\mbox.\egroup
  }{2010}]{benevenuto2010detecting}
Benevenuto, F.; Magno, G.; Rodrigues, T.; and Almeida, V.
\newblock 2010.
\newblock Detecting spammers on twitter.
\newblock In {\em Collaboration, electronic messaging, anti-abuse and spam
  conference (CEAS)}, volume~6, ~12.

\bibitem[\protect\citeauthoryear{Burke}{2017}]{lonewolf}
Burke, J.
\newblock 2017.
\newblock {The myth of the ‘lone wolf’ terrorist}.

\bibitem[\protect\citeauthoryear{Burnap and Williams}{2016}]{burnap2016us}
Burnap, P., and Williams, M.~L.
\newblock 2016.
\newblock Us and them: identifying cyber hate on twitter across multiple
  protected characteristics.
\newblock {\em EPJ Data Science} 5(1):11.

\bibitem[\protect\citeauthoryear{Cha \bgroup et al\mbox.\egroup
  }{2010}]{cha2010measuring}
Cha, M.; Haddadi, H.; Benevenuto, F.; and Gummadi, P.~K.
\newblock 2010.
\newblock Measuring user influence in twitter: The million follower fallacy.
\newblock {\em Icwsm} 10(10-17):30.

\bibitem[\protect\citeauthoryear{Chatzakou \bgroup et al\mbox.\egroup
  }{2017a}]{chatzakou2017hate}
Chatzakou, D.; Kourtellis, N.; Blackburn, J.; De~Cristofaro, E.; Stringhini,
  G.; and Vakali, A.
\newblock 2017a.
\newblock Hate is not binary: Studying abusive behavior of \#gamergate on
  twitter.
\newblock In {\em Proceedings of the 28th ACM Conference on Hypertext and
  Social Media}, HT '17.
\newblock ACM.

\bibitem[\protect\citeauthoryear{Chatzakou \bgroup et al\mbox.\egroup
  }{2017b}]{chatzakou2017mean}
Chatzakou, D.; Kourtellis, N.; Blackburn, J.; De~Cristofaro, E.; Stringhini,
  G.; and Vakali, A.
\newblock 2017b.
\newblock Mean birds: Detecting aggression and bullying on twitter.
\newblock {\em arXiv preprint arXiv:1702.06877}.

\bibitem[\protect\citeauthoryear{Chau and Xu}{2007}]{chau2007mining}
Chau, M., and Xu, J.
\newblock 2007.
\newblock Mining communities and their relationships in blogs: A study of
  online hate groups.
\newblock {\em International Journal of Human-Computer Studies}.

\bibitem[\protect\citeauthoryear{Chen, Conroy, and
  Rubin}{2015}]{chen2015misleading}
Chen, Y.; Conroy, N.~J.; and Rubin, V.~L.
\newblock 2015.
\newblock Misleading online content: recognizing clickbait as false news.
\newblock In {\em Proceedings of the 2015 ACM on Workshop on Multimodal
  Deception Detection},  15--19.

\bibitem[\protect\citeauthoryear{Davidson \bgroup et al\mbox.\egroup
  }{2017}]{davidson2017automated}
Davidson, T.; Warmsley, D.; Macy, M.; and Weber, I.
\newblock 2017.
\newblock Automated hate speech detection and the problem of offensive
  language.
\newblock {\em arXiv preprint arXiv:1703.04009}.

\bibitem[\protect\citeauthoryear{Dhingra \bgroup et al\mbox.\egroup
  }{2016}]{dhingra2016tweet2vec}
Dhingra, B.; Zhou, Z.; Fitzpatrick, D.; Muehl, M.; and Cohen, W.~W.
\newblock 2016.
\newblock Tweet2vec: Character-based distributed representations for social
  media.
\newblock {\em arXiv preprint arXiv:1605.03481}.

\bibitem[\protect\citeauthoryear{Djuric \bgroup et al\mbox.\egroup
  }{2015}]{djuric2015hate}
Djuric, N.; Zhou, J.; Morris, R.; Grbovic, M.; Radosavljevic, V.; and
  Bhamidipati, N.
\newblock 2015.
\newblock Hate speech detection with comment embeddings.
\newblock In {\em Proceedings of the 24th International Conference on World
  Wide Web},  29--30.

\bibitem[\protect\citeauthoryear{Fast, Chen, and
  Bernstein}{2016}]{fast2016empath}
Fast, E.; Chen, B.; and Bernstein, M.~S.
\newblock 2016.
\newblock Empath: Understanding topic signals in large-scale text.
\newblock In {\em Proceedings of the 2016 CHI Conference on Human Factors in
  Computing Systems},  4647--4657.
\newblock ACM.

\bibitem[\protect\citeauthoryear{Gerstenfeld, Grant, and
  Chiang}{2003}]{gerstenfeld2003hate}
Gerstenfeld, P.~B.; Grant, D.~R.; and Chiang, C.-P.
\newblock 2003.
\newblock Hate online: A content analysis of extremist internet sites.
\newblock {\em Analyses of social issues and public policy} 3(1):29--44.

\bibitem[\protect\citeauthoryear{Gitari \bgroup et al\mbox.\egroup
  }{2015}]{gitari2015lexicon}
Gitari, N.~D.; Zuping, Z.; Damien, H.; and Long, J.
\newblock 2015.
\newblock A lexicon-based approach for hate speech detection.
\newblock {\em International Journal of Multimedia and Ubiquitous Engineering}
  10(4):215--230.

\bibitem[\protect\citeauthoryear{Golub and Jackson}{2010}]{golub2010naive}
Golub, B., and Jackson, M.~O.
\newblock 2010.
\newblock Naive learning in social networks and the wisdom of crowds.
\newblock {\em American Economic Journal: Microeconomics} 2(1):112--149.

\bibitem[\protect\citeauthoryear{Greevy and
  Smeaton}{2004}]{greevy2004classifying}
Greevy, E., and Smeaton, A.~F.
\newblock 2004.
\newblock Classifying racist texts using a support vector machine.
\newblock In {\em Proceedings of the 27th annual international ACM SIGIR
  conference on Research and development in information retrieval},  468--469.
\newblock ACM.

\bibitem[\protect\citeauthoryear{Grover and
  Leskovec}{2016}]{grover2016node2vec}
Grover, A., and Leskovec, J.
\newblock 2016.
\newblock node2vec: Scalable feature learning for networks.
\newblock In {\em Proceedings of the 22nd ACM SIGKDD international conference
  on Knowledge discovery and data mining},  855--864.
\newblock ACM.

\bibitem[\protect\citeauthoryear{Guerra \bgroup et al\mbox.\egroup
  }{2017}]{guerra2017antagonism}
Guerra, P. H.~C.; Nalon, R.; Assunção, R.; and Jr., W.~M.
\newblock 2017.
\newblock Antagonism also flows through retweets: The impact of out-of-context
  quotes in opinion polarization analysis.
\newblock In {\em Eleventh International AAAI Conference on Weblogs and Social
  Media (ICWSM 2017)}.

\bibitem[\protect\citeauthoryear{Hamilton, Ying, and
  Leskovec}{2017a}]{hamilton2017inductive}
Hamilton, W.~L.; Ying, R.; and Leskovec, J.
\newblock 2017a.
\newblock Inductive representation learning on large graphs.
\newblock {\em arXiv preprint arXiv:1706.02216}.

\bibitem[\protect\citeauthoryear{Hamilton, Ying, and
  Leskovec}{2017b}]{hamilton2017representation}
Hamilton, W.~L.; Ying, R.; and Leskovec, J.
\newblock 2017b.
\newblock Representation learning on graphs: Methods and applications.
\newblock {\em arXiv preprint arXiv:1709.05584}.

\bibitem[\protect\citeauthoryear{{Hate Base}}{2017}]{hatebase}
{Hate Base}.
\newblock 2017.
\newblock {Hate Base}.

\bibitem[\protect\citeauthoryear{{Kadri, Thomas}}{2018}]{slate}
{Kadri, Thomas}.
\newblock 2018.
\newblock {Speech vs. Speakers}.

\bibitem[\protect\citeauthoryear{Kayes \bgroup et al\mbox.\egroup
  }{2015}]{kayes2015social}
Kayes, I.; Kourtellis, N.; Quercia, D.; Iamnitchi, A.; and Bonchi, F.
\newblock 2015.
\newblock The social world of content abusers in community question answering.
\newblock In {\em Proceedings of the 24th International Conference on World
  Wide Web},  570--580.
\newblock International World Wide Web Conferences Steering Committee.

\bibitem[\protect\citeauthoryear{Kwok and Wang}{2013}]{kwok2013locate}
Kwok, I., and Wang, Y.
\newblock 2013.
\newblock Locate the hate: Detecting tweets against blacks.
\newblock In {\em AAAI}.

\bibitem[\protect\citeauthoryear{Magu, Joshi, and
  Luo}{2017}]{magu2017detecting}
Magu, R.; Joshi, K.; and Luo, J.
\newblock 2017.
\newblock Detecting the hate code on social media.
\newblock {\em arXiv preprint arXiv:1703.05443}.

\bibitem[\protect\citeauthoryear{Meme}{2016}]{operationgoogle}
Meme, K.~Y.
\newblock 2016.
\newblock {Operation Google}.

\bibitem[\protect\citeauthoryear{Mikolov \bgroup et al\mbox.\egroup
  }{2013}]{mikolov2013efficient}
Mikolov, T.; Chen, K.; Corrado, G.; and Dean, J.
\newblock 2013.
\newblock Efficient estimation of word representations in vector space.
\newblock {\em arXiv preprint arXiv:1301.3781}.

\bibitem[\protect\citeauthoryear{Pennington, Socher, and
  Manning}{2014}]{pennington2014glove}
Pennington, J.; Socher, R.; and Manning, C.
\newblock 2014.
\newblock Glove: Global vectors for word representation.
\newblock In {\em Proceedings of the 2014 conference on empirical methods in
  natural language processing (EMNLP)},  1532--1543.

\bibitem[\protect\citeauthoryear{Rainie, Anderson, and
  Albright}{2017}]{rainie2017future}
Rainie, L.; Anderson, J.; and Albright, J.
\newblock 2017.
\newblock The future of free speech, trolls, anonymity and fake news online.
\newblock {\em Pew Research Center, March} 29.

\bibitem[\protect\citeauthoryear{Ribeiro \bgroup et al\mbox.\egroup
  }{2012}]{ribeiro2012sampling}
Ribeiro, B.; Wang, P.; Murai, F.; and Towsley, D.
\newblock 2012.
\newblock Sampling directed graphs with random walks.
\newblock In {\em INFOCOM, 2012 Proceedings IEEE}.
\newblock IEEE.

\bibitem[\protect\citeauthoryear{Ribeiro, Wang, and
  Towsley}{2010}]{ribeiro2010estimating}
Ribeiro, B.; Wang, P.; and Towsley, D.
\newblock 2010.
\newblock On estimating degree distributions of directed graphs through
  sampling.

\bibitem[\protect\citeauthoryear{Riloff \bgroup et al\mbox.\egroup
  }{2013}]{riloff2013sarcasm}
Riloff, E.; Qadir, A.; Surve, P.; De~Silva, L.; Gilbert, N.; and Huang, R.
\newblock 2013.
\newblock Sarcasm as contrast between a positive sentiment and negative
  situation.
\newblock In {\em EMNLP}, volume~13,  704--714.

\bibitem[\protect\citeauthoryear{Ross \bgroup et al\mbox.\egroup
  }{2017}]{ross2017measuring}
Ross, B.; Rist, M.; Carbonell, G.; Cabrera, B.; Kurowsky, N.; and Wojatzki, M.
\newblock 2017.
\newblock Measuring the reliability of hate speech annotations: The case of the
  european refugee crisis.
\newblock {\em arXiv preprint arXiv:1701.08118}.

\bibitem[\protect\citeauthoryear{Sabatini and
  Sarracino}{2017}]{sabatini2017online}
Sabatini, F., and Sarracino, F.
\newblock 2017.
\newblock Online networks and subjective well-being.
\newblock {\em Kyklos} 70(3):456--480.

\bibitem[\protect\citeauthoryear{Schmidt and Wiegand}{2017}]{schmidt2017survey}
Schmidt, A., and Wiegand, M.
\newblock 2017.
\newblock A survey on hate speech detection using natural language processing.
\newblock In {\em Proceedings of the Fifth International Workshop on Natural
  Language Processing for Social Media. Association for Computational
  Linguistics, Valencia, Spain},  1--10.

\bibitem[\protect\citeauthoryear{Shi \bgroup et al\mbox.\egroup
  }{2017}]{shi2017wisdom}
Shi, F.; Teplitskiy, M.; Duede, E.; and Evans, J.
\newblock 2017.
\newblock The wisdom of polarized crowds.
\newblock {\em arXiv preprint arXiv:1712.06414}.

\bibitem[\protect\citeauthoryear{Silva \bgroup et al\mbox.\egroup
  }{2016}]{silva2016analyzing}
Silva, L.~A.; Mondal, M.; Correa, D.; Benevenuto, F.; and Weber, I.
\newblock 2016.
\newblock Analyzing the targets of hate in online social media.
\newblock In {\em ICWSM},  687--690.

\bibitem[\protect\citeauthoryear{Stein}{1986}]{stein1986history}
Stein, E.
\newblock 1986.
\newblock History against free speech: The new german law against the"
  auschwitz" - and other - "lies".
\newblock {\em Michigan Law Review} 85(2):277--324.

\bibitem[\protect\citeauthoryear{{The Guardian}}{2017}]{youtubeboycott}
{The Guardian}.
\newblock 2017.
\newblock {Google's bad week: YouTube loses millions as advertising row reaches
  US}.

\bibitem[\protect\citeauthoryear{Twitter}{2017}]{twitterguidelines}
Twitter.
\newblock 2017.
\newblock {Hateful conduct policy}.

\bibitem[\protect\citeauthoryear{Van~Hee \bgroup et al\mbox.\egroup
  }{2015}]{van2015automatic}
Van~Hee, C.; Lefever, E.; Verhoeven, B.; Mennes, J.; Desmet, B.; De~Pauw, G.;
  Daelemans, W.; and Hoste, V.
\newblock 2015.
\newblock Automatic detection and prevention of cyberbullying.
\newblock In {\em International Conference on Human and Social Analytics (HUSO
  2015)},  13--18.
\newblock IARIA.

\bibitem[\protect\citeauthoryear{Viswanath \bgroup et al\mbox.\egroup
  }{2015}]{viswanath2015strength}
Viswanath, B.; Bashir, M.~A.; Zafar, M.~B.; Bouget, S.; Guha, S.; Gummadi,
  K.~P.; Kate, A.; and Mislove, A.
\newblock 2015.
\newblock Strength in numbers: Robust tamper detection in crowd computations.
\newblock In {\em Proceedings of the 2015 ACM on Conference on Online Social
  Networks},  113--124.
\newblock ACM.

\bibitem[\protect\citeauthoryear{Warner and
  Hirschberg}{2012}]{warner2012detecting}
Warner, W., and Hirschberg, J.
\newblock 2012.
\newblock Detecting hate speech on the world wide web.
\newblock In {\em Proceedings of the Second Workshop on Language in Social
  Media},  19--26.
\newblock Association for Computational Linguistics.

\bibitem[\protect\citeauthoryear{Waseem and Hovy}{2016}]{waseem2016hateful}
Waseem, Z., and Hovy, D.
\newblock 2016.
\newblock Hateful symbols or hateful people? predictive features for hate
  speech detection on twitter.
\newblock In {\em SRW @ HLT-NAACL},  88--93.

\bibitem[\protect\citeauthoryear{Waseem \bgroup et al\mbox.\egroup
  }{2017}]{waseem2017understanding}
Waseem, Z.; Chung, W. H.~K.; Hovy, D.; and Tetreault, J.
\newblock 2017.
\newblock Understanding abuse: A typology of abusive language detection
  subtasks.
\newblock In {\em Proceedings of the First Workshop on Abusive Language
  Online},  78--85.

\bibitem[\protect\citeauthoryear{Waseem}{2016}]{waseem2016you}
Waseem, Z.
\newblock 2016.
\newblock Are you a racist or am i seeing things? annotator influence on hate
  speech detection on twitter.
\newblock In {\em Proceedings of the 1st Workshop on Natural Language
  Processing and Computational Social Science},  138--142.

\end{thebibliography}

\end{small}

\end{document}